\documentclass{article}
\usepackage{graphicx}

\begin{document}

\begin{center}
Renormalization-Scale Invariance, Minimal Sensitivity, and the Inclusive Hadronic 
Decays of a 115 GeV Higgs Particle 
\end{center}

\begin{center}
F. A. Chishtie and V. Elias\\
Department of Applied Mathematics\\
The University of Western Ontario\\
London, Ontario   N6A 5B7\\
Canada
\end{center}

\begin{abstract}

Known perturbative expressions for the decay rates of 115 GeV Higgs particle into either 
two gluons or a $b\bar{b}$ pair are shown to exhibit renormalization-scale-($\mu$)-dependence that 
is largely removed via renormalization-group/Pad\'e-approximant estimates of these rates' 
next order contributions.  The extrema in $\mu$ characterizing both rates, as determined 
from fully-known orders of perturbation theory, are very nearly equal to corresponding 
$\mu$-insensitive rates obtained via estimation of their next order contributions, 
consistent with ``minimal-sensitivity'' expectations. 
\end{abstract}

Physical hadronic processes, such as measurable decay rates, are necessarily independent 
of the renormalization-scale parameter $\mu$ that characterises order-by-order calculations 
within perturbative QCD. Significant renormalization-scale dependence, however, {\it does} appear within QCD calculations 
that are truncated after only two or three orders of perturbation theory.  Since the 
Feynman-diagrammatic calculation of subsequent orders of QCD is very often prohibitively 
difficult, a common practise for decay rates is to identify $\mu$ with the mass of the decaying particle.  
Alternatively, one might either choose a value of $\mu$ for which the $\mu$-dependence of the known 
terms of the rate is locally flat, reflecting (hopefully) the renormalization-scale 
invariance of the all-orders calculation, or one might choose a value of $\mu$ for which 
the highest-order known term of the perturbative series is minimized, thereby forcing an 
(apparent) convergence of the known terms of the perturbative series. 

All such procedures 
for identifying $\mu$, however, are necessarily prescriptions for obtaining a prediction, and 
any discrepancies between predictions from differing procedures necessarily contributes to 
the theoretical uncertainty of such predictions. One purpose of the present work is to 
demonstrate how this (often overlooked) component of the theoretical uncertainty   
can be reduced via estimation of the next-order of the  perturbative series from 
which theoretical predictions are obtained. 

	In the present work, we consider the two gluon ($gg$) and $b\overline{b}$ decay modes of a Standard 
Model Higgs particle (H), as obtained from spontaneous electroweak symmetry breaking via a 
single complex scalar-field doublet.  We assume that $M_H$ = 115 GeV, as suggested by recent 
exciting results emerging from LEP \cite{AA}. The non-mass-suppressed contributions to the 
$H \rightarrow gg$ and $H \rightarrow b\overline{b}$ rates have been respectively calculated 
to two and three nonleading orders 
of perturbative QCD.  We show here that these calculated rates indeed exhibit 
renormalization-scale dependence, although this dependence is more pronounced for $H \rightarrow 
gg$ than for $H \rightarrow b\overline{b}$. We discuss how the next term in the perturbative series for both 
processes can be estimated via asymptotic Pade approximant methods developed by Samuel 
and collaborators \cite{AB,AC}. We demonstrate the success of such methods in predicting those 
next-order terms accessible via renormalization-group methods, and show how the incorporation 
of the estimated next-order term dramatically reduces the theoretical uncertainty associated 
with renormalization-scale dependence for both processes. We then obtain predictions for $H \rightarrow
gg$ and $H \rightarrow b\overline{b}$ decay rates for a 115 GeV Higgs particle for which those theoretical 
uncertainties associated with renormalization-scale dependence are virtually eliminated. 

The Higgs $\rightarrow$ gg decay rate, as extracted from correlation 
functions in the large $(2 M_t / M_H)^2$ limit \cite{AD}, may be expressed 
in the following form \cite{AE}

\begin{equation}
\Gamma_{H\rightarrow gg} = \frac{\sqrt{2} G_F M_H^3}{72\pi} S(\mu)
\end{equation}
where the renormalization-scale-sensitive factor in the decay rate (1)
is given by
\begin{eqnarray}
S(\mu) & = & S[x(\mu), L(\mu), T] \nonumber \\
& = & x^2 \left\{ 1 + x \left[ \left( \frac{215}{12} - \frac{23}{6} T
\right) + \frac{7}{2} L \right] \right. \nonumber \\
& + & x^2 \left[ 146.8912 - \frac{4903}{48} T 
+ \frac{529}{48} T^2 + \left( \frac{1445}{16} - \frac{161}{8} T \right) L 
+ \frac{147}{16} L^2 \right] \nonumber \\
& + & x^3 \left[ c_0 (T) + c_1 (T) L + c_2 (T) L^2 + c_3 (T) L^3 \right] \nonumber \\
& + &  \left. {\cal{O}} (x^4) \right\} 
\end{eqnarray}
with
\begin{equation}
x(\mu) \equiv \alpha_s^{(6)}(\mu) / \pi,
\end{equation}
\begin{equation}
L(\mu) \equiv \log (\mu^2 / m_t^2 (\mu)),
\end{equation}
\begin{equation}
T \equiv \log (M_H^2 / M_t^2)_{pole}.
\end{equation}
Equation (2) reflects the fact that the decay rate is known to 
three-loop order, and includes an appropriate parameterization of the subsequent unknown
four-loop-order contribution.

For the decay rate of a 115 GeV Higgs particle into two gluons, this
scale-sensitive factor is completely determined at $\mu = M_H$ by the
known central values of $\alpha_s^{(6)} (M_z) = 0.1185$ GeV 
(from which coupling-constant evolution can proceed to arbitarary $\mu$)
and $M_t = 174.3$ GeV \cite{AF}.  Expressed in terms of its three known leading orders,
this scale-sensitive factor (2) is given by
\begin{equation}
S(M_H)_{3-loop} = 0.00248 = (0.1148/\pi)^2 [1 + 0.657 + 0.201] .
\end{equation}
Because of its relatively large final term, the truncated perturbative
series within (6) appears likely to underestimate the true value of the
series sum.

The factor $S(\mu)$ satisfies the renormalization group (RG) equation \cite{AE},
\begin{equation}
0 = \mu^2\frac{dS}{d\mu^2} \left[ x(\mu), L(\mu), T \right] = 
\left[ 1 - 2\gamma_m^{(6)} (x) \right] \frac{ \partial S}{\partial L}
+ \beta^{(6)} (x) \frac{\partial S}{\partial x},
\end{equation}
a reflection of the renormalization-scale invariance of the physical
decay rate.  Perturbative expansions of the right-hand side of (7) can
be utilized to obtain order-by-order cancellation of scale-sensitivity
within $S(\mu)$.  Thus the value of $\mu$ chosen in (6) should, in
principle, not affect the rate obtained via (1).  However, truncation of the perturbation
series within (2) and (6) necessarily introduces a residual scale-dependent within the
rate.  In Figure 1, we have plotted the three-loop expression for $S(\mu)$ for $\mu$
over the entire $m_b (m_b) < \mu < M_t$ range of physical interest.  The three loop rate 
is manifestly renormalization-scale sensitive at $\mu$ = 115 GeV, exhibiting an increase 
with decreasing $\mu$ until local scale-insensitivity is obtained near $\mu$ = 49.5 GeV:
\begin{equation}
S(49.5 \; GeV)_{3-Loop} = 0.00254 = (0.1293/\pi)^2 [1+0.475 + 0.026].
\end{equation}
Note that the convergence of the series within (8) is much more evident 
than within (6), reinforcing the suggestion that (6) 
underestimates the full series sum at $\mu = M_H$.

In previous work \cite{AE}, we have demonstrated how asymptotic Pad\'e approximant methods 
can predict the four-loop coefficients $c_i (T)$ within (2), where $T$ is explicitly given 
by the renormalization-scale invariant logarithm (5):    
$T = 2 \log (115.0 / 174.3) = -0.8317$.  Thus, the perturbative series within (2) is of the form
\begin{equation}
S(\mu) / x^2 (\mu) = 1 + R_1 (L) x + R_2 (L) x^2 + R_3 (L) x^3 + ... ,
\end{equation}
where
\begin{equation}
R_1 (L) = 21.10 + \frac{7}{2} L
\end{equation}
\begin{equation}
R_2 (L) = 239.5 + 107.1 L + \frac{147}{16} L^2
\end{equation}
\begin{equation}
R_3 (L) = c_0 + c_1 L + c_2 L^2 + c_3 L^3
\end{equation}
The asymptotic Pad\'e-approximant prediction \cite{AG},
\begin{equation}
R_3 \approx \frac{2 R_2^3}{R_1^3 + R_1 R_2},
\end{equation}
as derived \cite{AH} from the anticipated error \cite{AC} of $[N|1]$ approximants in 
predicting the term $R_{N+2} x^{N+2}$ within the series (9), 
can be matched to the form of (12) by equating the four lowest moments of 
both versions of $R_3$:
\begin{equation}
w \equiv m_t^2 (\mu) / \mu^2, 
\end{equation}
\begin{eqnarray}
N_k & \equiv & (k+2) \int_0^1 dw \;w^{k+1} \left[ c_0 - c_1 \log w + c_2 \log^2
w - c_3 \log^3 w\right]\nonumber\\
& = & (k+2) \int_0^1 dw \; w^{k+1} \left[ \frac{ 2R_2^3 (L)}{R_1^3 (L) + 
R_1 (L) R_2 (L)} \right] , \; (L = -log(w)).
\end{eqnarray}
The final line of (15) is evaluated numerically via explicit
substitution of (10) and (11) into the integrand of (15), leading to the
following four equations:
\begin{equation}
N_{-1} = c_0 + c_1 + 2c_2 + 6c_3 = 4267
\end{equation}
\begin{equation}
N_0 = c_0 + c_1/2 + c_2/2 + 3c_3/4 = 2858
\end{equation}
\begin{equation}
N_1 = c_0 + c_1/3 + 2c_2/9 + 2c_3/9 = 2493
\end{equation}
\begin{equation}
N_2 = c_0 + c_1/4 + c_2/8 + 3c_3/32 = 2329
\end{equation}
with solutions
\begin{equation}
c_0 = 1900, \; \; c_1 = 1527, \; \; c_2 = 359.3, \; \; c_3 = 20.20
\end{equation}

The accuracy of these solutions is, in part, testable by comparison of
the predicted values for $\{c_1, c_2, c_3\}$ with the values one can
extract from the RG equation (7).  Given the
perturbative expansions of $n_f = 6$ RG functions
\begin{equation}
\beta^{(6)}(x) = -x^2 \left[ \frac{7}{4} + \frac{13}{8} x -
\frac{65}{128} x^2 + {\cal{O}}(x^3) \right]
\end{equation}
\begin{equation}
\gamma_m^{(6)} (x) = -x - \frac{27}{8} x^2 + {\cal{O}}(x^3)
\end{equation}
the coefficients $c_1$, $c_2$, and $c_3$ are easily obtained by the
requirement that the aggregate coefficients of $x^5$, $x^5 L$, and $x^5
L^2$ on the right-hand side of the RG equation (7) all vanish:
\begin{equation}
c_1^{RG} = 1540.4, \; \; c_2^{RG} = 364.83, \; \; c_3^{RG} = 21.4375 \;
.
\end{equation}
The striking agreement between the true values (23) and the estimated
values (20) suggests that the estimate in (20) of the RG-inaccessible
coefficient $c_0$ is comparably accurate.  One test of the internal
consistency and stability of the estimation procedure is to substitute
the known values (23) for $\{c_1, c_2, c_3\}$ into eqs. (16-19), and
then to solve each of these equations for $c_0$:

\begin{equation}
N_{-1} : c_0 = 1868
\end{equation}
\begin{equation}
N_0 : c_0 = 1890
\end{equation}
\begin{equation}
N_1 : c_0 = 1894
\end{equation}
\begin{equation}
N_2 : c_0 = 1896
\end{equation}
All four of these values are in good agreement with the estimate $c_0 = 1900$ 
[eq. (20)] obtained {\it without} the use of RG values for $\{c_1, c_2,
c_3\}$. Moreover, the estimate (24) also follows from
optimization $d\chi^2 / dc_0 = 0$ of the least-squares fit of (12) to (13)
over the UV domain of $\mu^2$, as discussed in \cite{AH}:
\begin{eqnarray}
\chi^2 [c_0] & = & \int_0^1 dw \left\{ \frac{2 R_2^3 (L)}{R_1^3 (L) + R_1 (L) R_2 (L)} \right. \nonumber \\
 & - & \left. c_0 + c_1^{RG} log(w) - c_2^{RG}log^2(w) + c_3^{RG}log^3(w) \right\}^2
\end{eqnarray} 

Thus, to estimate the four loop contribution to $S(\mu)$, we choose to incorporate
the estimate (24) for $c_0$ in conjunction with the known values (23)
for $\{c_1, c_2, c_3 \}$.  We also incorporate leading mass corrections \cite{AI}
to the rate, which can be expressed as a further contribution 
$\delta_m x^2 (\mu)$ to $S(\mu)$, where \cite{AE}
\begin{equation}
\delta_m = \frac{9}{16} \left[ (A_t + Re A_b)^2 + (Im A_b)^2 \right] -
1,
\end{equation}
\begin{equation}
A_t = 2 \left[ \tau_t + (\tau_t - 1) (sin^{-1} (\sqrt{\tau_t}))^2 \right]
/ \tau_t^2, \; \; \; \; \tau_t \equiv M_H^2 / 4 M_t^2
\end{equation}
\begin{eqnarray}
A_b = 2 \left[ \tau_b + (\tau_b - 1) f(\tau_b) \right] /
\tau_b^2,\nonumber \\
\tau_b \equiv M_H^2 / 4 m_b^2,\nonumber \\
f(\tau) = -\frac{1}{4} \left[ \log \frac{(1+\sqrt{1-1/\tau})}{(1 -
\sqrt{1-1/\tau})} - i \pi \right]^2 .
\end{eqnarray}
This correction to  $S(\mu)$ is quite small compared to the
contributions of the leading three-orders of perturbation theory, but
is comparable in size to the four loop term.  For example, using the
``physical'' mass values $M_t = 174.3$ GeV, $M_H = 115$ GeV, and $m_b =
4.20$ GeV we find that $\delta_m = -0.0597$, and that the mass-corrected
four-loop version of (6) is given by
\begin{equation}
S(M_H)_{4-loop + \delta_m} = 0.00245 = (0.1148/\pi)^2 [1 + 0.657 + 0.201 +
0.0373 \; (-0.0597)].
\end{equation}
In (32), the final number in parantheses is the mass-correction factor
$\delta_m$, preceded by the one-loop (unity), two-loop, three-loop and
(estimated) four-loop contributions to the rate in the $4M_t^2 / M_H^2 >> 1$ limit,
as obtained via (2), (23) and (24).  These series terms exhibit on
accelerated convergence over the four-orders of the perturbation series
within (32);  the ratio of successive terms (respectively 66\%, 31\% and
18.5\%) is itself seen to decrease as the order increases.
More importantly, the mass-corrected four-loop rate $S(\mu)$ is virtually
flat over a very broad range of $\mu$ inclusive of $\mu = M_H$, as is
evident from the ``4L'' curve displayed in Figure 2.  This curve
is seen to exhibit a very weak local maximum at
$\mu=110.5$ GeV which differs infinitesimally from the $S(M_H)$ rate of
(32);  {\it i.e.} $\left[ S(110.5\; GeV) / S(115 \; GeV) \right]_{4-
loop+\delta_m} = 1.00002$.  Had we not included the mass correction
factor $\delta_m$, this point of minimal sensitivity to 
renormalization-scale would have occurred at 74.5 GeV.

In Figure 2, we have also displayed the mass-corrected version of the 
(fully-known) three-loop (3L) expression for $S(\mu)$; for both the 3L 
and 4L curves in the Figure, the leading-order mass correction $\delta_m x^2(\mu)$ 
is included in  S($\mu$). The 3L curve clearly exhibits more renormalization-scale 
dependence than the 4L curve. However, the magnitude of the local maximum of the 3L curve 
(occurring at $\mu$ = 57.5 GeV) is in very good agreement with the plateau portion 
of the 4L curve, {\it i.e.} the estimated four-loop rate over the broad region of 
insensitivity to the renormalization-scale $\mu$. Such agreement of the 3L curve's 
extremum with the scale-insensitive next-order result is anticipated by 
arguments presented by Stevenson two decades ago \cite{AO}, in which the physical 
result to a given order of perturbation theory is identified with the point of 
local insensitivity to an (ultimately) unphysical parameter. The near 
equivalence of the 3L curve's extremum with the 4L curve's plateau supports the 
prescription of choosing a renomalization scale for which a perturbatively-obtained 
expression is "minimally sensitive" \cite{AO}, {\it i.e.} locally flat.
Thus, the factor $S(\mu)$ within the rate (1) is estimated to
four orders of perturbation theory by (32), corresponding to
$\Gamma_{H\rightarrow gg} = 2.71 \cdot 10^{-4}$ GeV if $M_H = 115$ GeV.
As is evident from Fig. 2, this estimate is insensitive to the magnitude of the renormalization-
scale parameter over a very broad domain of $\mu$ beginning at $\mu
\cong 20$ GeV and inclusive of $\mu = M_H$, in contrast to the known
three-loop expression for $S(\mu)$.

Within a Standard Model single-Higgs-doublet context, however, the process $H \rightarrow b\overline{b}$ 
is the dominant hadronic decay mode of a Higgs particle with mass 115 GeV.  
This decay rate has been obtained explicitly by Chetyrkin \cite{AJ}, and may be 
expressed in terms of a renormalization-scale dependent factor $R(\mu)$ as follows:
\begin{equation}
\Gamma(H \rightarrow b \overline{b}) = \frac{3G_F M_H}{4\pi \sqrt{2}} R(\mu)
\end{equation}
where, for $a \equiv \alpha_s^{(5)} (\mu) / \pi$, the scale-dependent factor $R(\mu)$ is 
proportional to the sum of $\Pi [a(\mu),\mu]$, the imaginary part of the scalar 
correlation function, and Higgs-mass-suppressed contributions $\Sigma [a(\mu), m_b(\mu)]$:
\begin{equation}
R(\mu) \equiv m_b^2 (\mu) \left[ \Pi[a(\mu),\mu] + \Sigma [a(\mu), m_b (\mu)] \right],
\end{equation}
\begin{eqnarray}
\Pi [a(\mu_,\mu] & = & 1 + a(\mu) \left[ \frac{17}{3} + 2 \log \left( \frac{\mu^2}{M_H^2} \right) \right]\nonumber \\
& + & a^2 (\mu) \left[ 29.14671 + \frac{263}{9} \log \left( \frac{\mu^2}{M_H^2} \right) + \frac{47}{12} \log^2 \left( \frac{\mu^2}{M_H^2} \right) \right] \nonumber \\
& + & a^3 (\mu) \left[ 41.75761 + 238.3806 \log \left( \frac{\mu^2}{M_H^2} \right) + 94.67590 \log^2 \left( \frac{\mu^2}{M_H^2} \right) \right.\nonumber \\
& + & \left. 7.615741 \log^3 \left( \frac{\mu^2}{M_H^2} \right) \right] \nonumber \\
& + & a^4 (\mu) \left[ d_0 + d_1 \log \left( \frac{\mu^2}{M_H^2} \right) + d_2 \log^2 \left( \frac{\mu^2}{M_H^2} \right) + d_3 log^3
\left( \frac{\mu^2}{M_H^2} \right) + d_4 \log^4 \left( \frac{\mu^2}{M_H^2} \right) \right] + ... \nonumber \\
\end{eqnarray}
\begin{equation}
\Sigma [a(\mu), m_b (\mu)] = \frac{m_b^2 (\mu)}{M_H^2} \left[ -6 - 40 \; a(\mu) - 87.72 \; a^2(\mu) + ... \right] .
\end{equation}
The factors of $M_H$ appearing in (33), (35) and (36) are pole masses that are independent 
of the renormalization scale parameter $\mu$.  In (35), we have included explicit 
parameterization of the ${\cal{O}}(a^4)$ contribution to $\Pi$. The coefficients $\{d_1, d_2, d_3, d_4\}$
can be extracted via the RG equation \cite{AK}
\begin{eqnarray}
\mu^2 \frac{d}{d\mu^2} \left[ m_b^2 (\mu) \Pi [a(\mu), \mu] \right] = 0 \nonumber \\
\Rightarrow \left[ \mu^2 \frac{\partial}{\partial \mu^2} + \beta^{(5)}[a]\frac{\partial}{\partial a} + 2 \gamma_m^{(5)} [a] \right] \Pi [a(\mu), \mu] = 0,
\end{eqnarray}
so as to yield the following values \cite{AK}:
\begin{equation}
d_1 = 791.5233, d_2 = 1114.695, d_3 = 260.0647, d_4 = 14.75550.
\end{equation}
However, the non-logarithmic coefficient $d_0$ is inaccessible via RG arguments at this order; 
its value can be obtained only by an explicit Feynman diagrammatic calculation which, as of 
the present time, has not been performed.

	Earlier work \cite{AE,AG,AK} has explored asymptotic Pade-approximant approaches toward the 
estimation of this $d_0$ term.  Given a field theoretical series of the form 
$1 + aR_1 + a^2R_2 + a^3R_3 + a^4R_4 + ...$, where $\{R_1, R_2, R_3\}$ are known and $R_4$ is unknown, 
the ${\cal{O}}(1|N)$ anticipated error of $[N|1]$ Pade-approximants in reproducing the series 
coefficient $R_{N+2}$ \cite{AC} can be utilized to predict that \cite{AG}
\begin{equation}
R_4^{pred.} = \frac{R_3^2[R_2^3+R_2 R_2 R_3 - 2R_1^3 R_3]}{R_2[2R_2^3 - R_1^3 R_3 - R_1^2 R_2^2]}.
\end{equation}
If, as in the pseudoscalar correlator  (35), the coefficient of $a^4$ is given by 
\begin{equation}
R_4 = d_0 + d_1 \log \left( \frac{\mu^2}{M_H^2} \right) + d_2 \log^2 \left( \frac{\mu^2}{M_H^2} \right) + d_3 \log^3
\left( \frac{\mu^2}{M_H^2} \right) + d_4 \log^4 \left( \frac{\mu^2}{M_H^2} \right),
\end{equation}
the coefficients $\{d_0, d_1, d_2, d_3, d_4\}$ can be predicted by an appropriate fitting of (39) 
to the form of (40).  The results of such fitting \cite{AK} are predicted values 
$[d_1= 745, d_2= 1180, d_3 = 253, d_4 = 15.4]$ within $\pm$6\% of the RG-determined values in (38), 
as well as a prediction that $d_0 = 64.2$. This prediction appears to be corroborated by the 
$(n_f = 5)$ estimate $d_0 = 62$ recently obtained by Broadhurst, Kataev, and Maxwell \cite{AL}, as well 
as an earlier estimate $(d_0 = 67.2)$ obtained via (39) entirely from nonlogarithmic terms \cite{AG}.

	In Figure 3 we have plotted $R(\mu)$, the renormalization-scale sensitive portion of the 
$H \rightarrow b\overline{b}$ decay rate for

\begin{description}
\item{1)} the case (4L) in which $\Pi$, as given by (35), is truncated after its fully-known ${\cal{O}}(a^3)$ contribution, and
\item{2)} the case (5L) in which the ${\cal{O}}(a^4)$ contributions to the correlator (35) are explicitly incorporated 
through use of the RG values (38) for the coefficients $\{d_1, d_2, d_3, d_4\}$ in conjunction with the estimate 
$d_0 = 64$ obtained in \cite{AK} via asymptotic Pade-approximant methods.
\end{description}

For both cases, the same mass-suppressed contribution $\Sigma [a(\mu), m_b(\mu)]$ is employed.  The final 
${\cal{O}}(a^2)$ term of this contribution, as given in (36), is generally smaller than the estimated  
${\cal{O}}(a^4)$ contribution to the pseudoscalar correlator (35).  In obtaining both Figure 3 curves, 
the evolution of $a(\mu)$ from the central empirical value \cite{AF} $\alpha_s(M_Z) = 0.1185$ is obtained via the 
known terms of the $n_f = 5$ $\beta$-function \cite{AM}:
\begin{equation}
\mu^2 \frac{da}{d\mu^2} = - \frac{23}{12} a^2 - \frac{29}{12} a^3 -
\frac{9769}{3456} a^4 - 18.85217 a^5.
\end{equation}
Similarly, the evolution of $m_b(\mu)$ from the assumed five-flavour threshold at 
$m_b(4.2 \; GeV) = 4.2 \; GeV$ is governed for both curves by the known terms of the $n_f = 5 \; \gamma_  
m$-function \cite{AN}:
\begin{eqnarray}
m_b(\mu) = 4.2 \; GeV \left[ \frac{a(\mu)}{a(4.2 \; GeV)}
\right]^{12/23} \frac{c[a(\mu)]}{c[a(4.2 \; GeV)]},\nonumber \\
c[a] = 1+1.17549a + 1.50071a^2 - 0.172486a^3.
\end{eqnarray}
A comparison of the two curves shows that inclusion of the estimated ${\cal{O}}(a^4)$ contribution to $\Pi$ 
leads to a significant flattening in the renormalization-scale dependence of the calculated 
$H \rightarrow b\overline{b}$ rate.  Once again, the extremum of $R_{4L}(\mu)$, 
the curve for which estimated ${\cal{O}}(a^4)$ contributions to $\Pi$ are {\it excluded}, corresponds to the 
broad $9.85 \; GeV^2$ plateau in $R_{5L}(\mu)$, the curve which does incorporate the estimated ${\cal{O}}(a^4)$ 
contributions to $\Pi$, consistent with the expectation that the point of 
minimal sensitivity to $\mu$ should correspond most closely to the true rate \cite{AO}.  
The extent of this plateau in $R_{5L}(\mu)$ can be ascertained by noting that 
$R_{5L}(\mu)$ is between 9.840 and 9.860 $GeV^2$ over a very broad domain of $\mu$ (22.5 GeV $ \leq \mu \leq$ 313 GeV).  
By contrast, the values of $R_{4L}(\mu)$ are within the same range only for a substantially smaller domain 
of $\mu$ (69.5 GeV $\leq \mu \leq$ 147 GeV).  Thus, it is evident that the inclusion of the 
asymptotic-Pade-approximant estimate of the next order contribution to the scalar correlator 
significantly reduces the renormalization-scale dependence of the predicted $H \rightarrow b \overline{b}$ 
decay rate. If we identify the true rate with the explicit plateau-region value 
$R_{5L}(\mu = 115 \; GeV) = 9.851 \; GeV^2$, the corresponding ($M_H = 115 \; GeV$) value for the decay rate 
is $\Gamma(H \rightarrow b\overline{b}) = 2.231 \cdot 10^{-3}\;  GeV$.

We conclude by noting for both $H \rightarrow gg$ and $H \rightarrow b
\overline{b}$ decay processes that if one considers {\it only the 
fully-known} terms of the appropriate perturbative series, our findings
support the ``minimal sensitivity'' prescription \cite{AO} in which
$\mu$ is assigned that value for which the rate is locally flat.  The
decay rates obtained from fully-known terms by this procedure are found
to be very nearly the same as those scale-insensitive rates we obtain
via estimation of the next-order contribution to the perturbative
series.

\bigskip

VE is grateful for support from the Natural Sciences and Engineering
Research Council of Canada.

\clearpage
\begin{figure}[hbt]
\centering
\includegraphics[scale=0.6, angle=270]{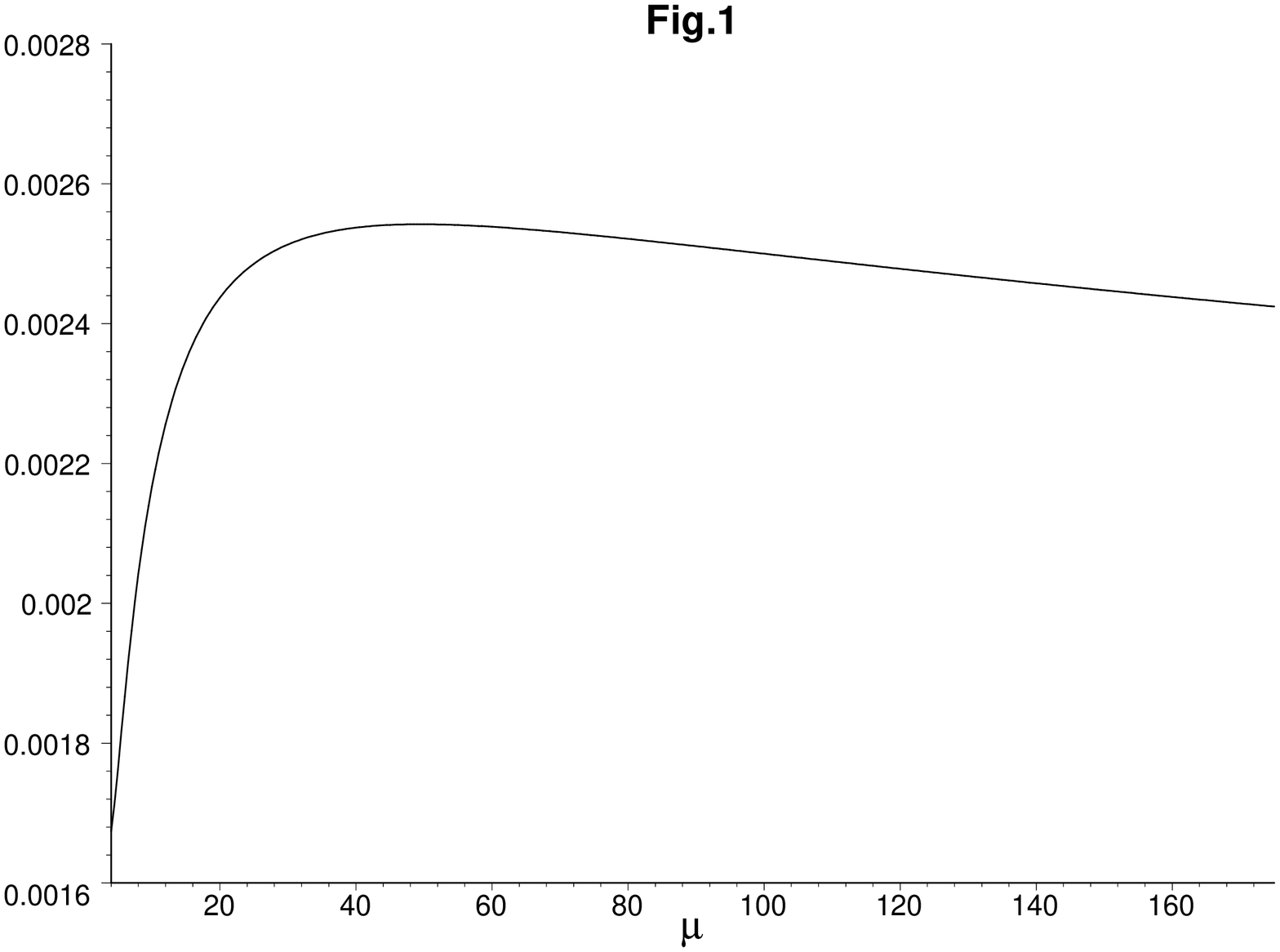}
\caption{The scale-sensitive portion $[S(\mu)]$ of the three-loop order Higgs   
$\rightarrow$ two gluon decay rate, as given by (1) and (2), is plotted for $M_t = 174.3$ GeV and 
$M_H = 115$ GeV, with the scale-dependent coupling evolving from  $\alpha_s(M_Z) = 0.1185$. 
The renormalization-scale parameter $\mu$ is in GeV; $S(\mu)$ is dimensionless. }
\label{fig1}
\end{figure} 

\clearpage
\begin{figure}[hbt]
\centering
\includegraphics[scale=0.6, angle=270]{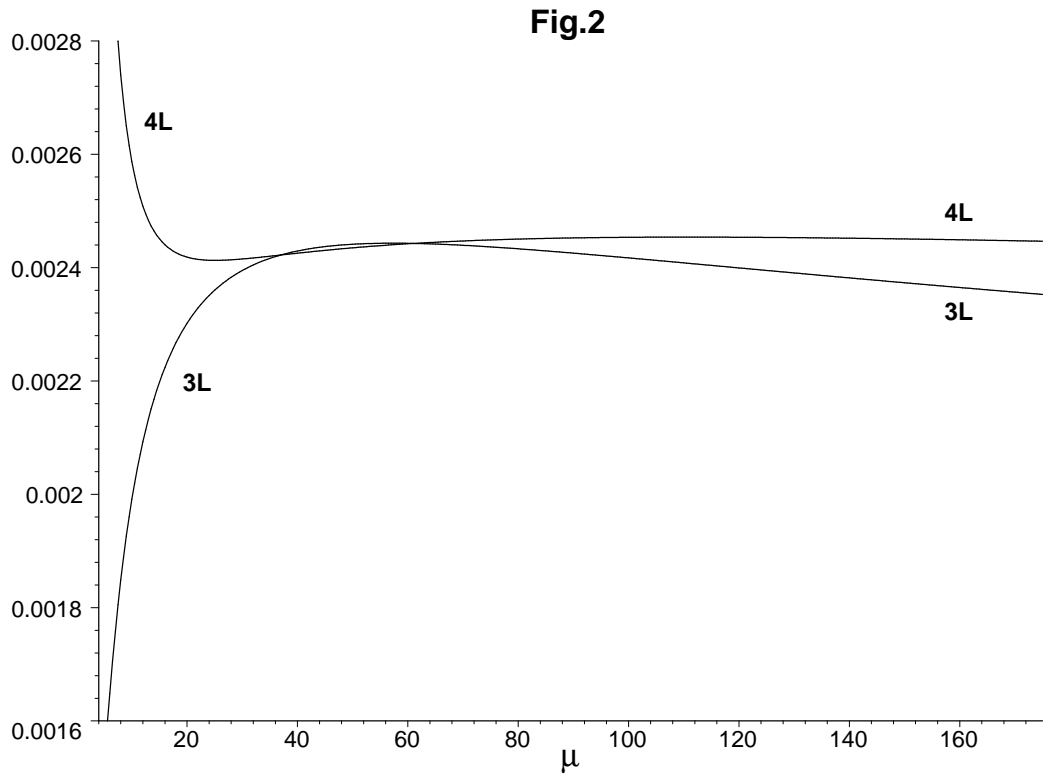}
\caption{The scale-sensitive portion $S(\mu)$ of the Higgs $\rightarrow$ two gluon decay rate, 
as obtained from fully-known contributions up to three-loop-order (3L), is compared 
to $S(\mu)$ as obtained from inclusion of its estimated next-order (4L) contributions. To 
isolate the effect of incorporating the next-order estimate, both expressions for $S(\mu)$ 
are inclusive of the mass correction  $\delta_m x^2(\mu)$, as obtained from (29-31) in the text, with 
$m_b = 4.2 \; GeV$, $M_t = 174.3 \; GeV$, $M_H = 115 \; GeV$, and the scale-dependent coupling evolving 
from  $\alpha_s(M_Z) = 0.1185$. The renormalization-scale parameter $\mu$ is in GeV; $S(\mu)$ is dimensionless.}
\label{fig2}
\end{figure}  

\clearpage
\begin{figure}[hbt]
\centering
\includegraphics[scale=0.6, angle=270]{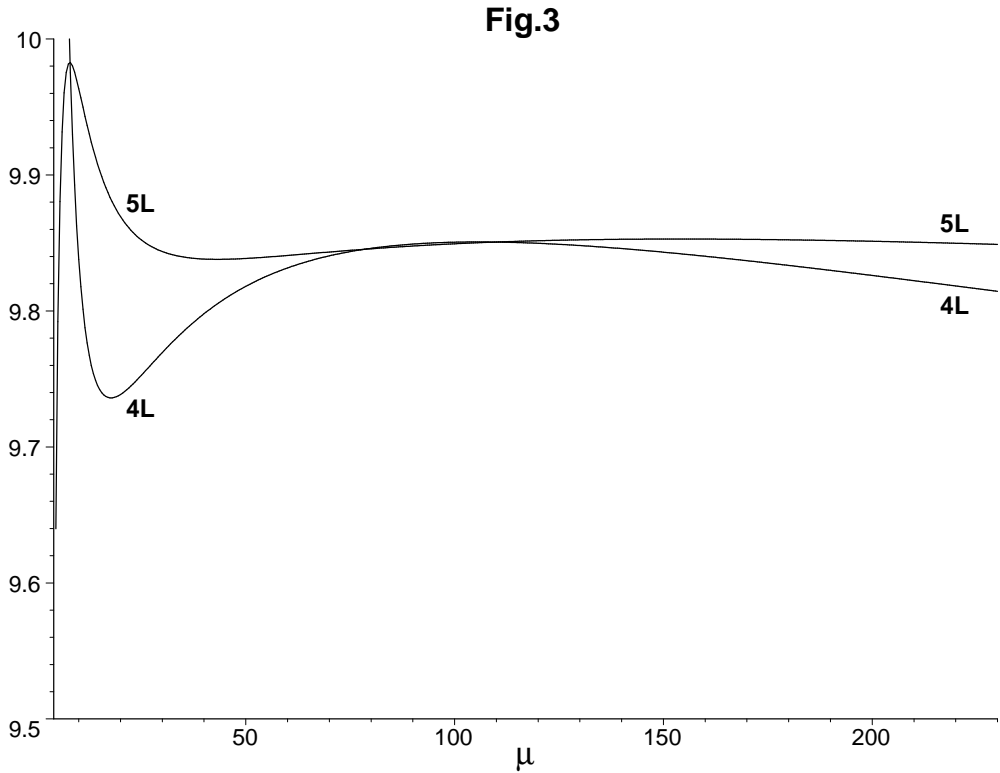}
\caption{The scale-sensitive portion $R(\mu)$ of the Higgs $\rightarrow b\bar{b}$ decay rate, as obtained from 
its fully-known four-loop-order (4L) contributions to the correlation function (35), is 
compared to $R(\mu)$ as obtained from inclusion of the scalar correlator's estimated next-order 
(5L) contributions. As before, $M_t = 174.3 \; GeV$, $M_H = 115 \; GeV$. The scale-dependent coupling 
and b-quark mass are evolved from  $\alpha_s(M_Z) = 0.1185$, $m_b(m_b)= 4.2 \; GeV$. The 
renormalization-scale parameter $\mu$ is in GeV, and the vertical axis is in GeV$^2$ units.}  
\label{fig3}
\end{figure}

\end{document}